# Fast Monte Carlo Simulation for Patient-specific CT/CBCT Imaging Dose Calculation

**Xun Jia, Hao Yan, Xuejun Gu, and Steve B. Jiang**

Center for Advanced Radiotherapy Technologies and Department of Radiation Oncology, University of California San Diego, La Jolla, CA 92037-0843, USA

E-mails: xujia@ucsd.edu, sbjiang@ucsd.edu

Recently, X-ray imaging dose from computed tomography (CT) or cone beam CT (CBCT) scans has become a serious concern. Patient-specific imaging dose calculation has been proposed for the purpose of dose management. While Monte Carlo (MC) dose calculation can be quite accurate for this purpose, it suffers from low computational efficiency. In response to this problem, we have successfully developed a MC dose calculation code, gCTD, on GPU architecture under the NVIDIA CUDA platform for fast and accurate estimation of the x-ray imaging dose received by a patient during a CT or CBCT scan. Techniques have been developed particularly for the GPU architecture to achieve high computational efficiency. Dose calculations using CBCT scanning geometry in a homogeneous water phantom and a heterogeneous Zubal head phantom have shown good agreement between gCTD and EGSnrc, indicating the accuracy of our code. In terms of improved efficiency, it is found that gCTD attains a speed-up of ~400 times in the homogeneous water phantom and ~76.6 times in the Zubal phantom compared to EGSnrc. As for absolute computation time, imaging dose calculation for the Zubal phantom can be accomplished in ~17 sec with the average relative standard deviation of 0.4%. Though our gCTD code has been developed and tested in the context of CBCT scans, with simple modification of geometry it can be used for assessing imaging dose in CT scans as well.

## 1. Introduction

Since its introduction in the 1970s, x-ray based computed tomography (CT) has been an important tool in medical imaging. It offers direct visualization of patient anatomy for a variety of purposes, including diagnostic or preventive screenings for certain diseases as well as image guidance in some therapy procedures. Due to its immense benefits, the usage of CT has increased dramatically over the last two decades (Smith-Bindman *et al.*, 2009). Along with these benefits, however, the potentially excessive x-ray imaging doses have recently become a serious concern. There has been an increasing desire to quantify patient-specific radiation dose from CT scans and to investigate the associated risks (Brenner, 2004; Brenner and Elliston, 2004; Bacher *et al.*, 2005; Hall and Brenner, 2008; Brix *et al.*, 2009; Alessio and Phillips, 2010; Li *et al.*, 2011b, a). This process, as illustrated in Fig. 1, will provide valuable information in patient medical records for dose monitoring and management. The first step toward realizing this is to accurately and promptly assess the radiation dose after every CT examination.

Among all the dose calculation methods, Monte Carlo (MC) simulation is commonly considered the most accurate method due to its capacity to faithfully describe the underlying physical interactions between radiation and matter. In addition, one can also accurately model the CT system, including the source location and scanner geometry, in MC to yield a high level of realism. Currently, general-purpose MC simulation packages, such as MCNP (Briesmeister, 1993), EGSnrc (Kawrakow, 2000), and PENELOPE (Baro *et al.*, 1995), have been utilized in many research works for CT dose calculation (Jarry *et al.*, 2003; DeMarco *et al.*, 2005; DeMarco *et al.*, 2007; Lee *et al.*, 2007; Li *et al.*, 2011a). Nonetheless, due to the extremely prolonged computation time, it is not very convenient to perform such an MC-based patient-specific dose calculation after every CT examination. Developing a fast and accurate MC CT dose calculation engine is therefore of clinical importance.

Another motivation for developing a fast CT dose calculation package is to facilitate radiotherapy treatments. In many image guided radiation therapy (IGRT) procedures, cone beam CT (CBCT) scans are performed on a daily basis before each treatment fraction for the purpose of patient positioning. These scans lead to a non-negligible amount of imaging doses on top of planed treatment dose, which may considerably alter the dose distribution received by a patient and hence the treatment outcomes (Ding and Coffey, 2009; Ding *et al.*, 2010). A CBCT dose calculation package will offer the possibility of routinely tracking the imaging dose, including it into treatment planning, and thus allowing for a better prediction of the total dose to cancerous target and critical organs (Alaei *et al.*, 2010).

Recently, graphics processing unit (GPU) has been increasingly utilized to speed up a number of computationally intensive tasks in medical physics (Samant *et al.*, 2008; Jacques *et al.*, 2008; Hissoiny *et al.*, 2009; Men *et al.*, 2009; Gu *et al.*, 2009; Jia *et al.*, 2010b; Gu *et al.*, 2010; Men *et al.*, 2010b; Men *et al.*, 2010a; Gu *et al.*, 2011a; Gu *et al.*, 2011b). In particular, GPU-based MC dose calculation packages have been developed for megavoltage energy range (Jia *et al.*, 2010a; Hissoiny *et al.*, 2011) and speed-up factors up to hundreds have been observed against conventional CPU-based dose engines. Though GPU can also be used for MC dose calculation in kilovoltage energy range, such a package is not yet available. Lately, a relevant package, MC-GPU (Badal and Badano, 2009), has been developed on a GPU platform to simulate x-ray transport using the MC method for the purpose of generating clinically-realistic x-ray projection images. With some necessary modifications, this package can be potentially utilized for dose calculation.

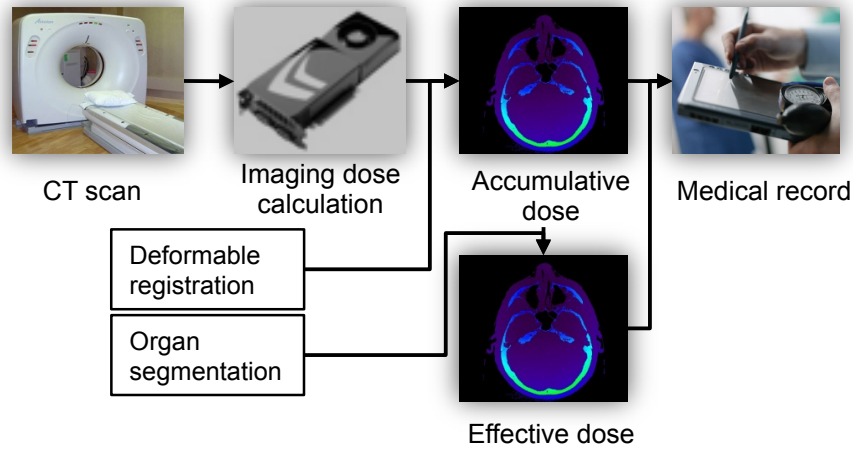

**Figure 1.** An illustration on the process of incorporating CT dose into patient's medical record based on the patient-specific CT dose calculation.

In this paper, we will present our recent work towards a high performance CT radiation dose calculation package, gCTD, on a GPU platform. Various techniques specifically tailored for GPU-based MC simulations have been employed to gain a high computational efficiency. Though the development of this package initially focuses and is tested on a CBCT context, with simple modifications of scanner geometry and source modeling, it can be easily adopted for other CT scanners. The rest of this paper is organized as follows. In Section 2, we describe the particle transport physics employed in gCTD. We will also present the algorithm structure and a number of key techniques in our implementation. Section 3 presents computation results and validations of our gCTD package. Finally, we conclude our paper in Section 4 and present some discussions.

## 2. Methods and Materials

### 2.1 gCTD Physics

In gCTD, the photon transport is handled by using Woodcock tracking method which significantly increases the simulation efficiency of the boundary crossing process (Woodcock *et al.*, 1965). In this method, the maximum total attenuation coefficient $\mu_{\max}(E)$ of the whole simulation volume at each energy $E$ is first computed. A virtual medium with a fictitious attenuation coefficient of $\mu_{\max}(E) - \mu_i(E)$ is then introduced to each voxel $i$ so as to compensate for the heterogeneity of the total attenuation coefficient among voxels. Because the geometry is now effectively homogeneous, the photons can be transported without the cumbersome ray tracing process and voxel boundary crossing checking. In particular, one can now sample the distance to the next interaction site, as if the photon is in a homogeneous medium with an attenuation coefficient $\mu_{\max}(E)$. Moving the photon to the next interaction site, an interaction type is sampled among Compton, Rayleigh, photoelectric absorption, and the fictitious interaction according to their attenuation coefficients. The photon state will be unchanged in a fictitious interaction, while physical processes corresponding to the other three interactions will take place in the cases of real interactions. Electron transport is not simulated in gCTD, but the energy of a secondary electron is locally deposited once it is generated in a Compton scattering event or a photoelectric event, since the range of those electrons is usually less than the voxel size (a few millimeter) in the keV energy range.

*2.2 GPU implementation*

A GPU card typically consists of a large number of scalar processor units, which allows us to perform massive parallel computation. For instance, in the context of MC simulation, it is very convenient to have each GPU thread be responsible for the transport of one source photon. Though the clock speed for each processor is lower than a typical CPU, the overall computational power is much higher due to the large amount of processors available on a single GPU. For example, the NVIDIA C2050 card used in this work is equipped with 448 processors with a clock speed of 1.15 GHz each. It also has 3 GB GDDR5 memory shared by all processor cores. Such a GPU card is designed and manufactured specifically for the purpose of scientific computing. It supports error correction codes to protect data from random errors occurred in data transfer and manipulation, ensuring computing accuracy and reliability. Our gCTD package is coded under the Compute Unified Device Architecture (CUDA) platform developed by NVIDIA (NVIDIA, 2010b). In this section, we will first present the overall simulation structure of our gCTD package and then discuss a few key techniques we employed.

*2.2.1 Overall structure*

The main structure of our code is shown in the left panel of Fig. 2. Once the simulation starts, the code is initialized with all the necessary data including the voxelized patient geometry, material properties, as well as all attenuation coefficient data. Random number seeds are also initialized at this step. All of these data are transferred to GPU memory during this step. Specifically, the voxelized phantom data and attenuation coefficients are stored in texture memory to allow for cached memory access and hardware supported linear interpolation, if necessary. Some constants repeatedly used during simulations, such as number of voxels in each dimension, are stored in constant memory of GPU. Global memory is used to store all other data, if not specified explicitly. After the initialization stage, simulation is performed in a batched fashion. We evenly divide the total number of histories into $N_b$ batches, *e.g.* $N_b = 10$. Inside each batch, a GPU kernel is launched to simulate a large number of source photons in parallel according the physical process discussed previously. Within the kernel, photons are first generated at the x-ray source and then transported till being absorbed or exiting the phantom. See the right panel of Fig. 2 for details. Dose depositions to voxels are also recorded during this process. After the simulations for all batches, statistical analysis is performed to obtain the average dose to each voxel and the corresponding statistical uncertainty. Finally, the program transfer data from GPU to CPU and outputs results before it terminates.

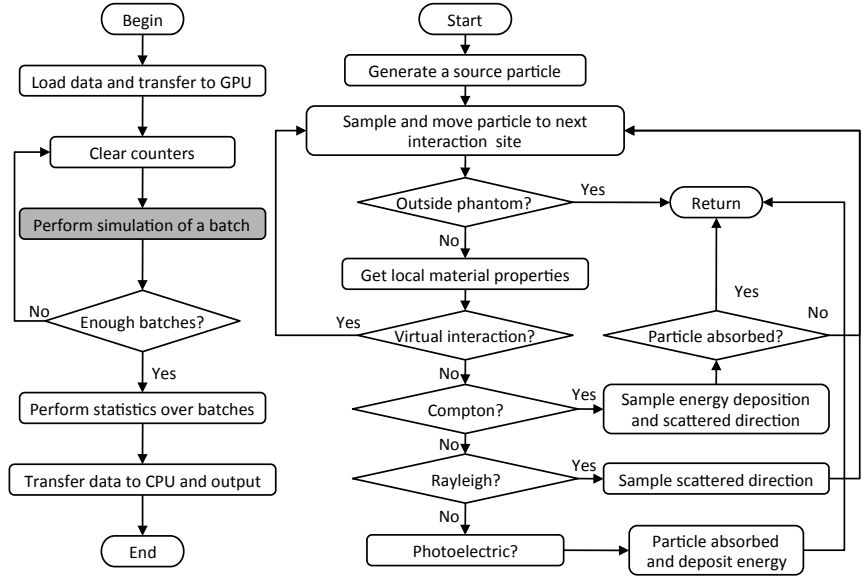

**Figure 2**. The flow chart of our gCTD MC simulation. Detailed steps of the batch simulation part (the shaded step in the left panel) are shown in the right panel.

### 2.2.2 Rayleigh scattering

In the case of Rayleigh scattering, the direction of the photon momentum $\boldsymbol{k}$ changes, while the photon energy remains the same. The scattering angle $\theta$ is sampled from Rayleigh differential cross section for a scattering molecule:

$$\frac{\mathrm{d}\sigma}{\mathrm{d}\Omega} = \frac{r_0^2}{2}(1 + \cos^2\theta)F^2(q^2), \tag{1}$$

where $q = 2k\sin\frac{\theta}{2}$ is the momentum transfer and $F^2(q^2)$ is the square of the form factor for a molecule. $k = \frac{E}{c}$ is the momentum of the incident photon with energy $E$. Note that $q$ is a function of $\theta$. Since $F^2(q^2) \to 0$ quickly as $q \to \infty$, the photon scattering direction is highly forward peaked. Under the independent atom approximation, $F^2(q^2) = \sum_i n_i F_i(q^2)$, where $n_i$ is the number of atom $i$ in the molecule and $F_i(q^2)$ is its form factor. In gCTD, we compute $F^2(q^2)$ using the atomic form factors $F_i(q^2)$ tabulated in the database of PENELOPE.

In many CPU-based MC simulation packages, such as EGS5 and PENELOPE, the sampling of a deflection angle $\theta$, or equivalently $\mu = \cos\theta$, is performed by using a generalized rejection method (Salvat *et al.*, 2009; Hirayama *et al.*, 2010). This method first samples a random number $\mu$ based on a probability density function (pdf) proportional to $F^2(q^2)$ and reject it with a probability proportional to $(1 + \mu^2)$. In practice, since $F^2(q^2)$ is tabulated in the computer memory, a certain kind of searching algorithm is required in this step. Though the table look-up is not a problem due to CPU's fast memory access and large cache space, this procedure is extremely computationally expensive on GPU, as latency for GPU memory access is very high and the uncontrolled search path among GPU threads may result in branching problems. To overcome this difficulty, we sample $\mu$ directly using an inverse transform method. Specifically, for a given energy $E$, we first numerically compute the pdf of $\mu$ as $p(\mu) = \frac{1}{Z}(1 + \mu^2)F^2[2k^2(1 - \mu)]$ at some discrete $\mu$ values equally spaced in $\mu \in [-1,1]$, where $Z$ is a normalization factor. The cumulative density function (cdf) $P(\mu) = \int_{-1}^{\mu} d\mu' \, p(\mu')$ is then numerically computed. With this cdf, we can easily compute its inverse function value $\mu = P^{-1}(\zeta)$ at a set of $\zeta$ values equally spaced in $\zeta \in [0,1]$, yielding a list of pairs $(\zeta_i, \mu_i)$. Finally, when the random variable $\mu$ is sampled, we can simply generate a random number $\zeta$ uniformly distributed in $[0,1]$ and $\mu = P^{-1}(\zeta)$ is obtained by a linear interpolation through the pair list of $(\zeta_i, \mu_i)$. Apparently, the $\mu$ value generated as such satisfies the desired distribution. Computation-wise, this method is extremely lightweight on GPU, since GPU supports hardware linear interpolation with a high efficiency. To handle the sampling for photons with various energy values encountered in the simulation, we first generate the pair $(\zeta_i, \mu_i)$ for a set of energy values $E_j$ ranging from 0 to the maximum energy, yielding $\mu_{i,j}$ defined on a 2d grid points $(\zeta_i, E_j)$. $\mu$ as a function of $\zeta$ and $E$ is denoted as R surface. During sampling, a 2d interpolation, again, using the GPU hardware at a random number $\zeta$ and the photon energy $E$ leads to a $\mu$ value. In practice, such R surfaces are pre-computed for each type of materials considered and are stored in GPU's texture memory to allow for the hardware interpolation.

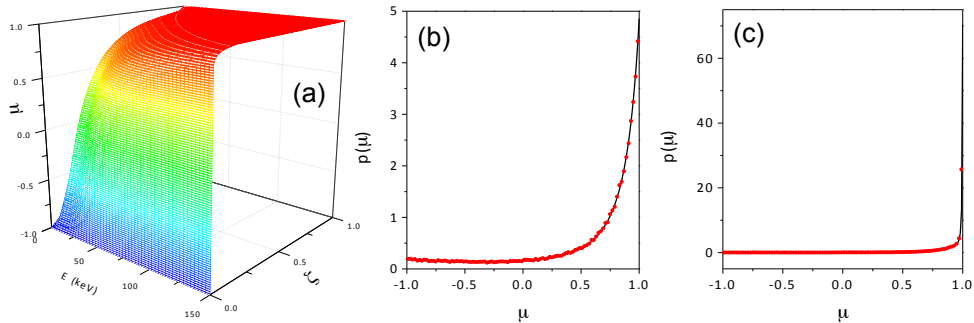

**Figure 3**. (a) The R surface for water generated for sampling the Rayleigh scattering angle. (b) and (c) are probability distribution functions of $\mu$ for water at 10 keV and 60 keV, respectively. Solid lines are from the theoretical calculation, and dots are estimated from $10^5$ samples.

To illustrate this method, the computed R surface is first plotted in Fig. 3(a) for water. For a given energy value, there is a large range of $\zeta$ such that $\mu \to 1$, indicating that the sampled scattering angles are forward peaked. As examples, Fig. 3(b) and (c) demonstrate the theoretical pdfs as well as those sampled from the R surface at 10keV and 60keV, respectively. Good agreements between them have been observed, indicating the effectiveness of our method.

### 2.2.3 Compton scattering

As for Compton scattering, the scattering angle $\theta$ is sampled from Compton differential cross section for a scattering molecule:

$$\frac{\mathrm{d}\sigma}{\mathrm{d}\Omega} = \frac{r_0^2}{2}\left(\frac{E'}{E}\right)^2 \left(\frac{E'}{E} + \frac{E}{E'} - \sin^2\theta\right) S(q), \tag{2}$$

where $E' = E/[1 + \frac{E}{m_e c^2}(1 - \cos\theta)]$ is the outgoing photon energy. $S(q)$ is called scattering function of a molecule, which is calculated here based on atomic scattering functions tabulated in the database of PENELOPE. The rest of the right hand side in Eq. (2) is known as Klein-Nishina differential cross section describing the Compton interaction of a photon with a free electron. The presence of $S(q)$ is due to the binding effect of the electron in an atom, leading to the vanishing probability of a photon being scattered in the forward direction. Again, conventional CPU-based MC packages samples the scattering angles by the generalized rejection method. In our implementation of gCTD, a C surface generated in a similar manner to what is described in the previous subsection is used. The C surfaces for all materials are pre-generated and are stored on GPU's texture memory. Fig. 4(a) depicts the C surface for water medium, while good agreements between the theoretical and the computed pdfs are found in Fig. 4(b) and (c).

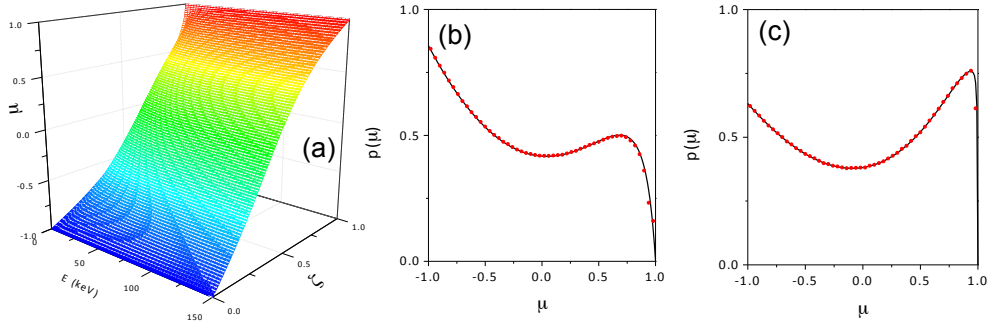

**Figure 4.** (a) The C surface for water generated for sampling Compton scattering angle. (b) and (c) are probability distribution functions of $\mu$ for water at 10 keV and 60 keV, respectively. Solid lines are from theoretical calculation, and dots are estimated from $10^6$ samples.

Yet, the differential cross section in Eq. (2) is based on the Waller-Hartree approximation, where only the binding effect is considered but not Doppler broadening effect. It is known that the Waller-Hartree approximation is less accurate than the uncorrected Klein-Nishina differential cross section in terms of dose calculation in the very low energy range (Salvat *et al.*, 2009). In gCTD, there is also a function available that samples the scattering angles from the simple Klein-Nishina differential cross section. Users have the option to select one of the two sampling methods.

### 2.2.4 Fluence map

X-ray source modeling is of crucial importance for dose calculation accuracy. There are two main components in source modeling. First of all, the particle fluence when exiting from the source is not homogeneous. For instance, Fig. 5(a) shows a measured image from an air scan in a Varian OBI CBCT system (Varian Medical Systems, Inc., Palo Alto, CA, USA) equipped with a full-fan bowtie filter, indicating a strong modulation of the particle fluence along the patient lateral direction. To model this effect, we use an air scan image as the desired fluence map and generate particle fluence accordingly. Specifically, if we divide the fluence map into a total number of $N_f$ small beamlets and label them in a certain order by an index $I = 1,2, \dots N_f$, the associated fluence map intensity $f_I$ represents the relative probability that a source photon comes from the beamlet $I$. The goal of sampling a photon following this fluence map can be achieved by first sampling a beamlet index $I$ according to the relative probably determined by $f_I$ and then sampling the particle inside this beamlet uniformly. For simplicity, let us assume that $f_I \neq 0$ for all beamlets $I$ considered. To ensure simulation efficiency, we utilize the so called Metropolis-Hastings sampling algorithm (Hastings, 1970). The key steps of this algorithm are illustrated in Algorithm A1. In this algorithm, the beamlet index generated for the last particle $I_{last}$ is stored, used, and updated each time a new beamlet index is generated.

---

**Algorithm  A1:**

Initialize $I_{last}$ with an arbitrary beamlet index in $\{1,2, \dots, N_f\}$.

Do the following steps each time a particle is generated:

1.  Generate a trial beamlet $J$ uniformly from $\{1,2, \dots, N_f\}$;
2.  Generate a random number $\zeta$ uniformly distributed in $[0,1]$;
3.  If $\zeta < f_J / f_{I_{last}}$, set $I = J$; otherwise set $I = I_{last}$.
4.  Generate a particle within the beamlet $I$ uniformly.
5.  Set  $I_{last} = I$.

---

It has been proven that such an algorithm is able to generate a sequence of beamlet indices, which follow the distribution governed by $f_I$, given that this sequence is long enough. Note that each time a new beamlet index is generated, only one memory access to the beamlet intensity $f_J$ is needed, which ensures the computational efficiency by avoiding frequent visits to the slow GPU memory. In practice, since we are performing parallel computation, each GPU thread is initialized with its own $I_{last}$ generated by a CPU random number at the initialization stage.

To demonstrate the convergence of this algorithm, we record the particles generated according to the measured air scan image shown in Fig. 5(a) and the resulting photon fluence with $10^8$ particles is depicted in Fig. 5(b). These two fluence maps are visually similar except for an obvious level of noise in the simulated map. To quantify this similarity, we compute the error $e = \left\| p_f - p_f^* \right\|$, where $p_f$ and $p_f^*$ are vectors composed of the probability at each beamlet for the simulated and measured fluence maps, respectively. As shown in Fig. 5(c), this error monotonically and quickly decreases, as the particle number increases.

5    *2.2.5 Energy spectrum*

Photons coming from an x-ray tube are not monoenergetic. Therefore, the source particle energy has to be generated according to an energy spectrum for accurate dose calculation. A conceivable way of taking this spectrum into the simulation is to randomly sample the energy for each source photon according to the energy spectrum. Yet, a large number of photons are simulated simultaneously on GPU and the computation time among them varies due to their different energies. As a consequence, those GPU threads with photons of shorter simulation time will have to wait for others with longer simulation time, which reduces the overall computational efficiency. To resolve this issue, we evenly divide the entire energy spectrum into a set of intervals of width 1 keV each. The total number of photons to be simulated is first distributed to each energy interval according to the spectrum. Simulation is then performed for each interval sequentially with the particle energy uniformly distributed inside the interval. This strategy ensures that, at any moment of the simulation, all GPU threads are dealing with source particles of similar energies, mitigating the efficiency loss due to the variation of simulation time between GPU threads.

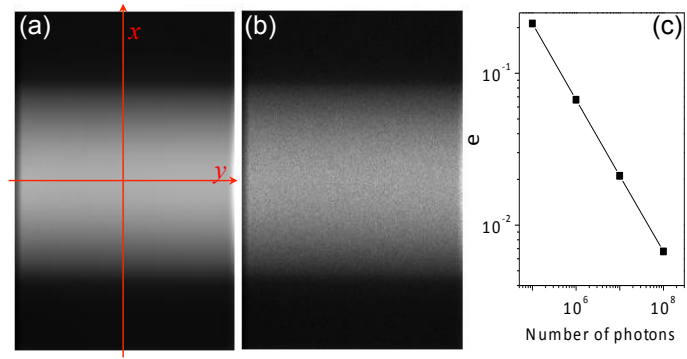

**Figure 5.** A measured fluence map for a CBCT source with a full-fan bowtie filter is shown in (a). The simulated fluence map with $10^8$ particles is shown in (b). (c) is the dependence of the relative error $e$ on the particle number simulated.

*2.2.6 Source trajectory*

25

The x-ray source is not at a static position during the scanning process. In a CBCT scan, for example, it moves around the patient along a circular trajectory. To account for this in our simulation, we first parameterize the motion trajectory as $\boldsymbol{r}(t) = (x(t), y(t), z(t))$. $t \in [0, T]$ parameterizes the time during which the scan is performed. For instance, the functions $x(t) = R \sin \omega t$, $y(t) = R \cos \omega t$, $z(t) = 0$ and $t = [0, 2\pi/\omega]$ correspond to a scan in which the source moves at a constant angular speed $\omega$ around a patient for $2\pi$ along a circle of a radius $R$. Once the trajectory is specified, each source photon is generated with the parameter $t$ chosen at random in the interval $[0, T]$ uniformly, yielding the source location $\boldsymbol{r}$ for the photon.

*2.2.7 Other issues*

There are a few other issues of importance for the accuracy and efficiency of a MC simulation. (1) Single-precision floating point data type is used throughout gCTD to represent rational numbers instead of double-precision that is used widely in many MC simulation packages. Apparent degradation of accuracy was not observed in validation studies. (2) We use a pseudo-random number generator provided by a library CURAND (NVIDIA, 2010a), which offers a light-weighted GPU function that produces simple and efficient generation of high-quality pseudo-random numbers using XORWOW algorithm (Marsaglia, 2003). The period of such a generator is about $2^{192}$ and the quality of the random numbers has been tested using the TestU01 "Crush" framework of tests (L'Ecuyer and Simard, 2007). (3) In gCTD, we use linear interpolation for all the attenuation coefficient data of physical interactions. This linear interpolation can be achieved by GPU hardware via the so called texture memory. Since the interpolation of attenuation coefficient data is a frequently performed task during a MC simulation, the use of hardware supported linear interpolation enhances the overall program efficiency considerably.

## 3. Results

In this section, we provide dose calculation results in one homogenous water phantom and one Zubal head-and-neck (HN) phantom to test our gCTD package. Doses calculated using gCTD are compared with those computed using EGSnrc (Kawrakow, 2000). A high level of agreement between them will clearly demonstrate the accuracy achieved with our gCTD system. Meanwhile, the computation time is recorded and compared to demonstrate the gain of computational efficiency. For the hardware used in this section, the GPU results are obtained on an NVIDIA Tesla C2050 card, while the desktop computer on which the EGSnrc code is executed is equipped with a 2.27 GHz Intel Xeon processor and 4GB memory.

*3.1 Phantom studies*

We first test our gCTD on a water phantom of a dimension $20.0 \times 20.0 \times 20.0$ cm$^3$ and the voxel size is set to be $0.4 \times 0.4 \times 0.4$ cm$^3$. For the testing purpose, the x-ray source is set to be static with a source-to-isocenter distance SID = 100.0 cm and the x-ray impinges normally to the phantom surface. Field size is chosen to be $26.7 \times 20.0$ cm$^2$ at the isocenter level, corresponding to an x-ray projection size of $40.0 \times 30.0$ cm$^2$ at the imager level with a source-to-imager distance of 150.0 cm. The x-ray source attains an energy spectrum of a 125 kVp beam from a tungsten target and 2mm Al filter (Boone and Seibert, 1997). X-ray source fluence after a full-fan bowtie filter shown in Fig. 5(a) is used. The photon absorption energy is set to be 1 keV in our simulation. A total number of $2.5 \times 10^8$ source photons are simulated in gCTD, which are evenly divided into 10 batches for the purpose of calculating statistical uncertainties. As for the calculation in EGSnrc, the source particle number is also $2.5 \times 10^8$.

The calculated dose distributions are presented in Fig. 6, where the results are normalized to the maximum dose in the phantom. Fig. 6(a) depicts the percentage depth dose curve along the beam central axis, whereas (b) and (c) are lateral profiles along two axes defined in Fig. 5(a) at $z = 3.0$ cm depth. Note the lateral profiles along these two directions are different due to the application of an x-ray bow-tie filter. The error bars represent the level of two times statistical uncertainty of the results from our gCTD code. The error bars corresponding to the EGSnrc results are of similar sizes and not drawn for the purpose of clarity. These figures clearly demonstrate good agreements between the gCTD results and the EGSnrc results.

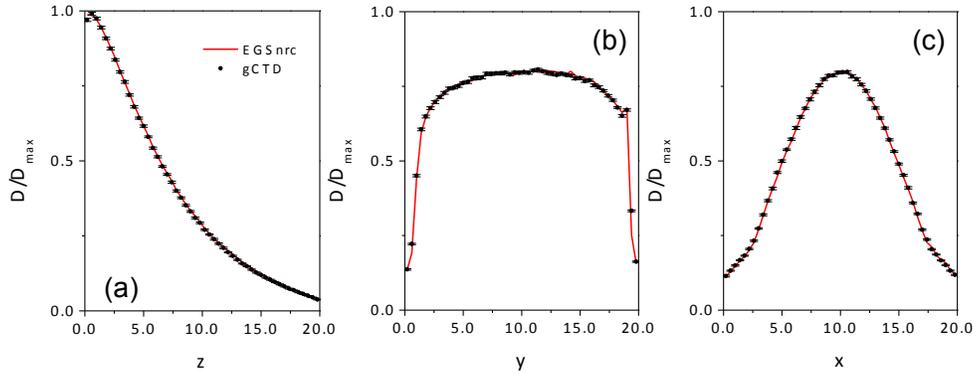

**Figure 6.** Percentage depth dose curve (a) and dose profiles ((b) and x (c)) at 3cm depth along two lateral directions (defined in Fig. 5(a)) in a water phantom.

To demonstrate the precision of our simulations, we calculate the uncertainty $s$ at each voxel normalized by the maximum dose $D_{\max}$. We further average the relative uncertainty $\sigma = s/D_{\max}$ over the high dose region where the local dose $D$ exceeds 20% of its maximum value $D_{\max}$ inside the phantom, yielding average relative uncertainty $\bar{\sigma}$. As indicated in Table 1, $\bar{\sigma}$ is found to be less than 1% for this phantom in both gCTD and EGSnrc.

We have also quantified the agreement between the gCTD results and the EGSnrc results using $\gamma$-index evaluation (Low *et al.*, 1998; Gu *et al.*, 2011b). Specifically, we compute the $\gamma$-index value $\gamma(\boldsymbol{r}) = \min_{\boldsymbol{r}'} \sqrt{\frac{\left(D_{EGSnrc}(\boldsymbol{r}) - D_{gCTD}(\boldsymbol{r}')\right)^2}{\Delta D^2} + \frac{(\boldsymbol{r} - \boldsymbol{r}')^2}{\Delta r^2}}$ for each voxel $\boldsymbol{r}$, where $D_{EGSnrc}$ and $D_{gCTD}$ are doses obtained from the two codes, respectively. $\Delta D$ and $\Delta r$ are parameters in the evaluation and are chosen as $\Delta D = 0.02 D_{\max}$ and $\Delta r = 2$mm in this work. It is understood that the voxel $\boldsymbol{r}$ passes the $\gamma$-index test, if $\gamma(\boldsymbol{r}) < 1$. Finally, the passing rate of this test $P_\gamma$ is computed over the region where $D > 0.2 D_{\max}$, which is simply the quotient of the number of voxels which pass the test and the total number of voxels inside this region. A passing rate $P_\gamma = 98.9\%$ is observed in this water phantom case, which quantitatively indicates the good agreement between the results from EGSnrc and from gCTD.

Finally, we evaluate the computational efficiency. The total computation time $T$ in both packages are recorded in our simulation. For our gCTD code, the time used to initialize simulation, such as transferring data from CPU to GPU, is also included. To make the comparison fair, for each of the two codes, we compute the simulation efficiency defined as $\epsilon = 1/\bar{\sigma}^2 T$. It is expected that such a metric is a good indication of the level of precision one can achieve within a certain computational time. We further use the ratio $\epsilon_{gCTD}/\epsilon_{EGSnrc}$ to characterize the gain of computation efficiency of our code against EGSnrc. It is found out that our gCTD considerably increases the simulation efficiency, by a factor of 414, in this water phantom case.

**Table 1**. Average relative uncertainty ($\bar{\sigma}$), $\gamma$ evaluation passing rate ($P_\gamma$), absolute computation time ($T$), and efficiency ($\epsilon$) for gCTD and EGSnrc in the water phantom and the Zubal phantom.

| Code | # of Histories | Phantom | $\bar{\sigma}$ (%) | $P_\gamma$ (%) | $T$ (sec) | $\epsilon_{gCTD}/\epsilon_{EGSnrc}$ |
|---|---|---|---|---|---|---|
| gCTD | $2.5\times10^8$ | Water | 0.64 | 98.9 | 6.95 | 414 |
| EGSnrc | $2.5\times10^8$ | Water | 0.70 | | 2454 | |
| gCTD | $2.5\times10^8$ | Zubal | 0.44 | 98.2 | 17.7 | 76.6 |
| EGSnrc | $2.5\times10^8$ | Zubal | 0.18 | | 8100 | |

*3.2 Zubal Phantom*

To further test our gCTD code in a heterogeneous medium, we have performed dose calculation for a Zubal head-and-neck (HN) phantom (Zubal *et al.*, 1994) in a CBCT scan. The x-ray source is the same as the one used in the previous water phantom case and it moves around the patient in a $200°$ angular as indicated by the arrow in Fig 7. The phantom body is described by a voxelized 3D image with a dimension of $128\times128\times60$ and the voxel size is $0.4\times0.4\times0.4$ cm. Based on the density, the material type of each voxel in the phantom is assigned to air, tissue, or bone. With a total of $2.5\times10^8$ source photons simulated, the imaging dose distribution calculated by our gCTD package is displayed in the top row of Fig. 7 overlayed on top of the density image of the phantom. The x-rays mainly deposit their energy in bony structures due to the high photoelectric attenuation coefficients of the bone relative to other materials. We have also performed an equivalent simulation using EGSnrc. The kinetic cutoff energy is set to 125 keV for electrons to avoid unnecessary simulation of electron transport at this low energy range. A very similar dose distribution has been observed in EGSnrc results. Specifically, the dose profiles along three major axes are plotted in Fig. 7, indicating a good agreement between the EGSnrc results and our results. Quantitatively, a very high $\gamma$ test passing rate $P_\gamma = 98.2\%$ is observed. As for the simulation efficiency, it is found that the efficiency of gCTD is about 76.6 times higher than that of EGSnrc.

Compared to the homogeneous water phantom case, the speedup factors drops considerably. This can be ascribed to the different photon transport mechanisms employed in the two packages. In gCTD, Woodcock transport is used, where a fictitious medium is introduced to make the phantom effectively homogeneous. The photon transport is performed in a homogeneous phantom with an attenuation coefficient equal to the largest physical attenuation coefficient among all voxels. Though cumbersome voxel boundary crossing is not necessary anymore with this technique, it introduces a large number of fictitious interactions where no real physical scattering takes place. It is this fact that limits the simulation efficiency of gCTD in heterogeneous phantoms. On the other hand, EGSnrc transport photons with explicit voxel boundary crossing. Its efficiency is hence not affected by the heterogeneity.

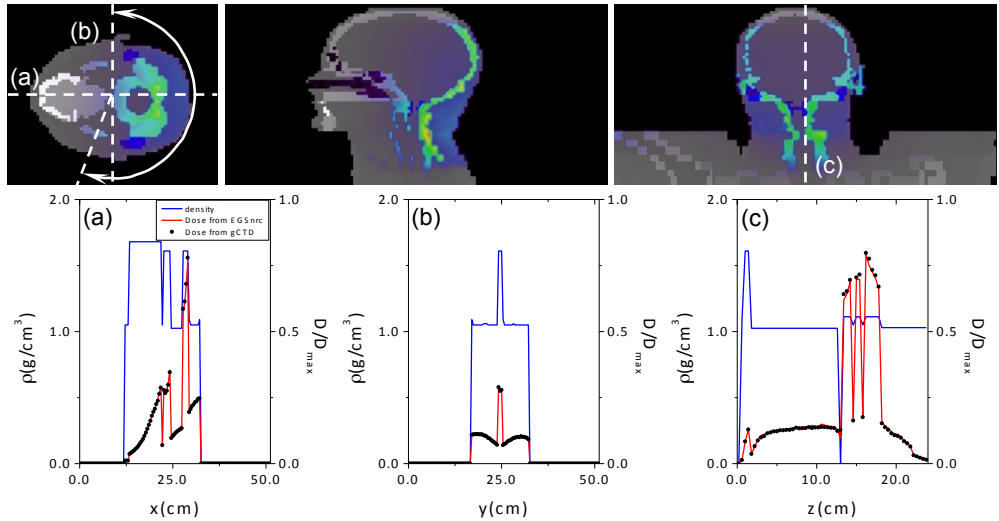

**Figure 7.** Top: imaging dose distribution in the Zubal phantom from a full fan CBCT scan. Bottom: Density plot and dose profiles along the three axes (a)~(c), respectively.

## 4. Discussion and Conclusions

In this paper, we have successfully developed a MC dose calculation package, gCTD, on a GPU architecture under NVIDIA CUDA platform for the purpose of quickly and accurately estimating x-ray imaging dose received by a patient from a CT/CBCT scan. Various techniques specifically tailored for the GPU architecture have been developed, resulting in a high computational efficiency. Dose calculations in a homogeneous water phantom and in a heterogeneous HN Zubal phantom have shown good agreements between gCTD and EGSnrc, indicating the accuracy of our code. In terms of efficiency, it is found that gCTD attains a speed-up of ~400 times in a homogeneous water phantom and ~76.6 times in a heterogeneous phantom compared to EGSnrc. In particular, imaging dose calculation for a HN phantom can be accomplished in ~17 sec with the average relative standard deviation of 0.4%. Yet, the efficiency comparison between gCTD and EGSnrc is not completely fair as the two packages employ quite different particle transport physics. We only emphasize here that the imaging dose calculation can be achieved with sufficient accuracy and precision in about tens of seconds with our package, which will greatly facilitate the studies involving patient-specific imaging dose calculation. Moreover, although our gCTD code is developed and tested in the context of CBCT scans, with simple modifications regarding scanning geometry, it can be readily applied to assess radiation dose in CT scans as well.

On the other hand, though it is now possible to perform MC dose calculation for a CBCT scan in ~17 sec, there is still room for improvement in terms of efficiency. On the software side, it is expected that variance reduction techniques will be very helpful. Currently, no variance reduction technique is yet implemented in gCTD and hence the convergence rate suffers from the stochastic nature of the particle transport process. Integration of variance reduction techniques, such as particle splitting and track repeating, should yield a further efficiency boost. On the hardware side, multi-GPU can be used as a powerful platform to further improve the efficiency. On such a platform, all the particle histories simulated can be distributed among all the GPUs, which then execute simultaneously without interfering with each other. Therefore, a roughly linear scalability of the computation efficiency can be achieved with respect to the number of GPUs. It has been reported recently that this linear scalability holds on a dual-GPU system (Hissoiny *et al.*, 2011) and on a 4-GPU system (Jia *et al.*, 2011) for megavoltage MC dose calculation. All of these strategies that can potentially improve the performance will be investigated in the future release of gCTD. With these strategies, it will become possible to assess patient imaging dose in near real time. Clinical introduction of such a package will greatly facilitate tracking CT/CBCT dose delivered to a patient and imaging dose management.

As discussed while studying the Zubal phantom case, heterogeneity in the phantom impacts on the dose calculation efficiency duo to the application of the Woodcock transport method. In the Zubal phantom case, the highest density found in bone is 2.75 g/cm$^3$, already very high in a typical patient case. Therefore, this phantom case represents a large population of real patient cases in terms of calculation time. Yet, larger amount of heterogeneity may still exist in real patient cases, for example, if metal pieces or metal artifacts are present. In these circumstances, the computation time is expected to be further prolonged.

Meanwhile, the feasibility of real-time CT dose monitoring depends not only on the efficiency of the MC simulation, but also on many other issues. For instance, before launching gCTD, the 3D CT dataset of a patient needs to be converted into a voxelized phantom. Deformable registration between the current CT image and a template CT image is necessary for the calculation of accumulated CT dose. Moreover, it is more clinically relevant to estimate dose to radiosensitive organs than to calculate dose to each voxel, which then requires auto-segmentation of relevant organs from a patient's CT images. Although gCTD itself does not imply the accomplishment of the entire process of patient-specific CT dose monitoring and management, as shown in Fig. 1, its development constitutes a solid step towards this by solving the low efficiency problem for dose estimation. It is beyond the scope of this paper to develop high efficiency tools to achieve other key components such as auto-segmentation, which will be research topics in future.

## 5  Acknowledgements

This work is supported in part by the University of California Lab Fees Research Program, the Master Research Agreement from Varian Medical Systems, Inc., and the Early Career Award from the Thrasher Research Fund. We would like to thank Dr. Frédéric Tessier from National Research Council Canada for their help regarding the use of EGSnrc and Miriam Graf for carefully proofreading the manuscript. We would also like to thank NVIDIA for providing GPU cards for this project.